\documentclass[twocolumn]{aa}
\usepackage{graphicx}
\newcommand{\COJ}[2]{\mbox{CO$\,J={#1} - {#2}$}}

\newcommand{\J}[2]{\mbox{$J={#1} - {#2}$}}
\newcommand{\JJ}[2]{\mbox{${#1} - {#2}$}}
\newcommand{\LOJ}[2]{\mbox{L$'_{\rm{CO}}({#1} - {#2})$}}
\newcommand{\Msolar}{\mbox{$M_{\odot}\,$}}
\newcommand{\Lsolar}{\mbox{$L_{\odot}\,$}}
\newcommand{\XCO}{\mbox{$X_{\rm{CO}}$}}              

\newcommand{\hours}{\mbox{$^{h}$}}

\newcommand{\mins}{\mbox{$^{m}$}}
\newcommand{\psec}{\mbox{$\stackrel{s}{\textstyle .}$}}

\newcommand{\degs}{\mbox{$^{o}$}}

\newcommand{\parcsec}{\mbox{$\stackrel{\prime\prime}{\textstyle .}$}}
\newcommand{\arcsecs}{\mbox{$^{\prime\prime}$}}

\newcommand{\arcmins}{\mbox{$^{\prime}$}}

\begin{document}
   \title{Detection of \COJ{1}{0} in the $z=3.79$ Radio Galaxy 4C\,60.07}

%   \subtitle{}

   \author{T.R.~Greve 
          \inst{1}
          \and
          R.J.~Ivison
          \inst{2}
          \and 
          P.P.~Papadopoulos
          \inst{3,4}
          \fnmsep
          }

   \offprints{tgreve@roe.ac.uk}

   \institute{Institute for Astronomy, University of Edinburgh,
              Blackford Hill, Edinburgh EH9 3HJ, United Kingdom\\
              \email{tgreve@roe.ac.uk}
         \and
             UK ATC, Royal Observatory, Blackford Hill, Edinburgh EH9 3HJ, United Kingdom\\
             \email{rji@roe.ac.uk}
	 \and 
	     Department of Physics \& Astronomy, University College London, Gower Street, London WC1E 6BT, United Kingdom\\
        \email{pp@star.ucl.ac.uk}
	\and
	     Sterrewacht Leiden,  P. O. Box 9513,  2300 RA Leiden, The Netherlands 
             }

   \date{Received December 2, 2003; accepted February 5, 2004}
 %  \date{}

   \abstract{We report on the detection of the lowest \COJ{1}{0} transition
in the powerful high-redshift radio galaxy 4C\,60.07 at $z$ = 3.79. 
The CO emission is distributed in two spatially and
kinematically distinct components as was previously known from the observations
of the higher excitation \COJ{4}{3} line. 
The total molecular gas mass in 4C\,60.07 inferred from the \COJ{1}{0} emission is
$M(\mbox{H}_2)\simeq 1.3\times 10^{11}\,\Msolar$, sufficient to fuel the inferred star-formation rate of 
$\sim 1600\,\Msolar\,$yr$^{-1}$ for $10^8$\,yrs. 
From our high-resolution \COJ{1}{0} VLA maps we find the dynamical mass
of 4C\,60.07 to be comparable
to that of a giant elliptical at the present time. A significant
fraction of the mass is in the form of molecular gas 
suggesting that 4C\,60.07 is in an early state of its evolution. 
The merging nature of 4C\,60.07 along with its large dynamical mass
imply that this system is a giant elliptical caught in its formative stages.
   }

   \maketitle
%
%________________________________________________________________

\section{Introduction}
High-redshift radio galaxies (HzRGs) are amongst the most luminous 
objects known, and are believed to serve as tracers of the peaks of
the primordial density field around which giant elliptical galaxies and
clusters of galaxies form (Kauffmann 1996; West et al.~1994). 
In the radio, HzRGs typically display a double-lobe
morphology and large radio luminosities ($P_{178\,MHz} \sim 5 \times 10^{35}$\,erg\,s$^{-1}$\,Hz$^{-1}$),
indicating a highly active black hole. 

Recently, evidence
has been mounting that HzRGs are massive starburst galaxies.
This has come about from sub-millimetre detections of a number of HzRGs, implying large
rest-frame far-IR luminosities ($L_{FIR} \simeq 10^{13}\,\Lsolar$) 
powered by intense star formation ($SFR \simeq 1000-2000\,\Msolar$\,yr$^{-1}$ -
Dunlop et al.~1994; Hughes et al.~1997; Ivison et al.~1998; Archibald et 
al.~2001). In a recent SCUBA survey of seven HzRGs and their surroundings,
Stevens et al.~(2003) not only found the star formation in the radio galaxies themselves
to be extended on several tens of kilo-parsec scales but also found one or 
more previously undetected submm 
sources in the vicinity ($50-250\,$kpc) of more than half of the targeted objects. 
It is difficult to see how the Active Galactic Nucleus (AGN) could power
the far-IR luminosity on $\ga 10$\,kpc scales, and a massive starburst seems
to be the natural explanation.
Indeed, adequate "fuel" for such large star formation rates has been found in the 
four HzRGs which have been detected in CO to date (Papadopoulos et al.~2000; 
De Breuck et al.~2003a; De Breuck et al.~2003b). 
These observations revealed the presence of massive ($\sim 10^{11}\,\Msolar$) 
reservoirs of molecular gas, enough to fuel a $\sim 1000\,\Msolar$\,yr$^{-1}$
starburst for $\sim 10^8$\,yr, and in in half of the cases the CO emission was found to be
extended on tens of kpc scales (Papadopoulos et al.~2000; De Breuck et al.~2003a).
Similar large molecular gas masses distributed in clumps on tens of kilo-parsec scales 
has been found in a number of QSOs at high redshifts (Carilli et al.~2002a; Carilli et al.~2002b).
In general, HzRGs have the advantage over quasars that they are not
gravitationally lensed since they are usually
selected on the basis of extended lobe-emission whereas quasars
are often found to be lensed. Furthermore, HzRGs are known to be associated with 
giant ellipticals in the local Universe (McLure \& Dunlop 2000).

In this paper we present high-resolution observations of the \COJ{1}{0}
emission from 4C\,60.07 at $z=3.788$ using the Very Large Array\footnote{The Very Large Array (VLA) is operated by
the National Radio Observatory, which is a facility of the National Science
Foundation, operated under cooperative agreement by Associated Universities,
Inc.}. 
Throughout we have assumed
$H_0=65$\,km\,s$^{-1}$\,Mpc$^{-1}$, $\Omega_M=0.3$ and $\Omega_\Lambda=0.7$. In
this cosmology the luminosity distance of 4C\,60.07 is 36.2\,Gpc and
$1\arcsecs$ corresponds to 7.7\,kpc.

\section{4C\,60.07}
\begin{figure}
   \centering
   \hspace*{-0.3cm}\includegraphics[width=1.0\hsize,angle=0]{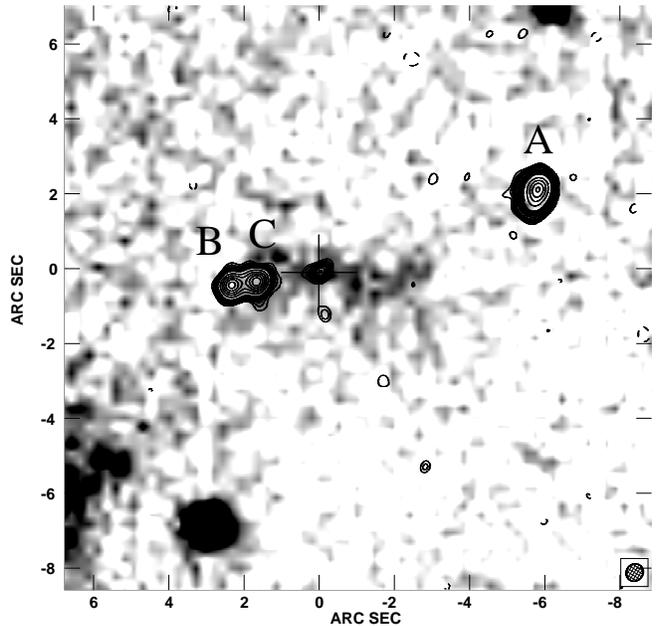}
      \caption{6\,cm VLA map of 4C\,60.07 overlaid on an I-band image
obtained with the William Herschel Telescope. The radio hot spots 
are denoted by A, B and C. The cross marks the position of
the radio core: R.A.(J2000): 05\hours12\mins55\psec147, DEC.(J2000): +60\degs30\arcmins51\parcsec0.
The insert in the bottom right corner shows the synthesized beam.}
         \label{fig:4c60-cband}
   \end{figure}

4C\,60.07 is an ultra-steep-spectrum (USS) radio galaxy at redshift of
$z=3.788$ (Chambers et al.~1996; R\"ottgering et al.~1997). It displays a Fanaroff-Riley II (FR II)
edge-brightened double-radio morphology (Fanaroff \& Riley 1974).
The radio morphology of 4C\,60.07 is seen in Figure \ref{fig:4c60-cband} which shows a VLA C-band (6\,cm) 
archive image of 4C\,60.07. The system consists of two main bright hot spots 
separated by about $9\arcsecs$. The south-eastern component is further comprised
of two components (B and C). Continuum emission is also seen from the radio core which
is located $\sim2\arcsecs$ west-northwest of the C-component.

The \COJ{4}{3} line and 1.25\,mm continuum emission from 4C\,60.07 have been imaged 
using the IRAM Plateau de Bure Interferometer (Papadopoulos et al.~2000).
The CO emission was found to emerge 
from two kinematically
distinct components separated in velocity space by $\sim 700 \rm{\, km\,
s}^{-1}$. The component with the narrowest line profile ($FWHM \sim 150 \rm{\,
km\, s}^{-1}$) is offset by $V\simeq-224$\,km\,s$^{-1}$ from the systemic velocity
corresponding to $z=3.788$ and spatially coincident with the position of the
radio core. The broader component ($FWHM \geq 550 \rm{\, km\, s}^{-1}$)
peaks $\sim 7 \arcsecs$ ($\sim 30\rm{\, kpc}$) west of the radio core, and is offset by 
$\simeq 483$\,km\,s$^{-1}$ from the systemic velocity. Such large offsets between the
redshift of the optical emission lines and the CO emission has been observed
in several other high redshift systems, and is commonly attributed to the optical
lines originating in strong outflows and winds (Guilloteau et al.~1999; Cox et al.~2002).
Strong gravitational lensing can be ruled out as
the origin of the double source and large apparent luminosity since not only do
the two components have different line widths but they are also offset in
velocity with respect to each other. Furthermore, there are no indications from
observations in the optical and radio suggesting that 4C\,60.07 might be
lensed.  
%----------------------------------------------------------- 
%______________________________________________________________

\begin{figure}
   \centering
   \includegraphics[width=1.0\hsize,angle=0]{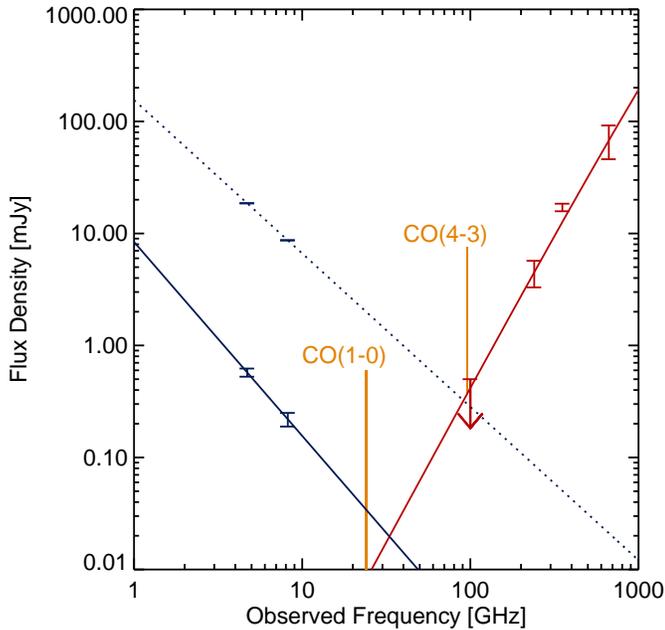}
      \caption{The submm/far-IR spectral energy distribution (SED) of 4C\,60.07 (red line) along with the
radio spectra of the A-component (blue dotted line) and the radio core (blue solid line). 
The submm points are taken from SCUBA observations by Archibald et al.~(2001) while the IRAM 30m 
Telescope data-points at 252\,GHz (1.25\,mm) and 100\,GHz (3\,mm) are from 
Papadopoulos et al.~(2000). The 
radio points are measured from C and X band data of 4C\,60.07 from the VLA Archive.
The positions and strengths of the \COJ{1}{0} and \J{4}{3} lines 
relative to the SED are shown in yellow (based
on this paper and Papadopoulos et al.~(2000)).}
         \label{fig:4c60-sed}
   \end{figure}

In order to get a handle on the non-thermal contribution to the submm and CO fluxes, 
we used C and X band images from the VLA Archive to measure the radio fluxes of the
various components in 4C\,60.07 at 6\,cm and 3.6\,cm. 
In Figure \ref{fig:4c60-sed} we have plotted the submm/far-IR spectral energy distribution (SED) of 4C\,60.07 
along with the radio spectra of the A-component and the radio core.
The A-component (the dotted line in Figure \ref{fig:4c60-sed}) is the brightest component 
in the radio and furthermore has the shallowest
spectral slope ($\alpha_{6\,cm}^{3.6\,cm} = -1.4$) which means it provides a strict upper limit
on the non-thermal flux at submm wavelengths. 
The 1.25\,mm continuum emission from 4C\,60.07 is likely to be thermal in origin, 
i.e.~from warm dust, since non-thermal processes are unable to account for the observed
flux, see Figure 2. Furthermore, the 1.25\,mm continuum emission is offset by
$\sim 4\arcsecs$ to the west from the radio core position, and does not appear
to be associated with the non-thermal emission. 

The radio core has a spectral slope of $\alpha_{6\,cm}^{3.6\,cm} = -1.7$. Extrapolating
to the frequency of the \COJ{1}{0} line ($\sim 24$\,GHz)
we find that the radio continuum is expected to contribute a non-neglible 
$S_{24\,\rm{GHz}} = 0.03\,$mJy. In comparison, the contribution from the 
thermal dust spectrum seems completely neglible, see Figure \ref{fig:4c60-sed}.
%
%----------------------------------------------------------- S_vib
%______________________________________________________________

\section{Observations}
\begin{figure}
   \centering
   \includegraphics[width=1.0\hsize,angle=0]{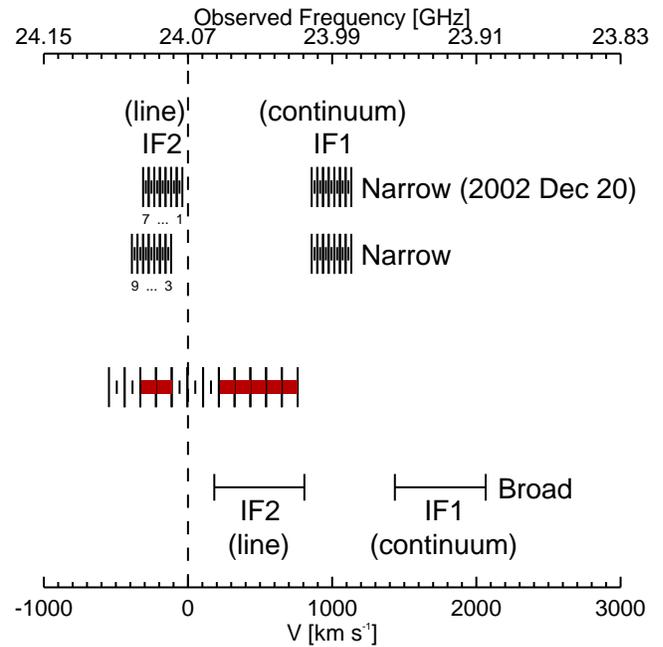}
      \caption{Velocity and frequency coverage of the IF set-up used in our VLA \COJ{1}{0} observations. The dashed line
outlines the systemic velocity corresponding to $z=3.788$. In red is shown
the velocity-channels of the IRAM PdBI observations of Papadopoulos et al.~(2000);
channels in which \COJ{4}{3} emission was detected are marked in thick red.}
         \label{fig:IF-setup}
   \end{figure}

The \COJ{1}{0} line ($\nu_{rest} = 115.27$\,GHz) from 4C\,60.07 
is redshifted into the VLA's K-band (1.3-cm) receivers. Since 
non-thermal continuum emission from the radio core was expected, we used two IF
pairs (right- and left-hand circular polarisation) centred on the two
kinematically distinct emission-line components, each pair set up such that one
IF pair was centred on the line and the other pair was offset from this by
100\,MHz to measure the continuum. The IF set-up for both the broad and narrow component is 
detailed in Figure \ref{fig:IF-setup}. The broad-line component was observed in continuum mode with a
50\,MHz-wide IF centred at 24.035\,GHz, which corresponds to a velocity
coverage of 624\,km\,s$^{-1}$. Not only does this not properly cover the broad component
but, since the observations were done in continuum mode, no
information on the line shape was available, and one therefore has to 
rely on the IRAM \COJ{4}{3} observations to infer a line width for the broad component.
The narrow component was observed in
spectral-line mode with a 7-channel IF centered at 24.095\,GHz, each channel
being 3.125\,MHz wide. This was to avoid under-resolving the narrow line
in velocity space. However, the line datasets from the 2002 December 20 and
2003 March 09 are centred at 24.089\,GHz, corresponding to a shift of two
channels, in order to get a better spectral coverage at the low-frequency side
of the line. Thus nine channels (denoted channels 1,..,9 in Figure \ref{fig:IF-setup})
covered the line, corresponding to a velocity
coverage of 350\,km\,s$^{-1}$. 4C\,60.07 was observed in the VLA's CD, D and C
configurations (see Table \ref{tbl-1}). The D-configuration data, however, 
turned out not to have the spatial resolution required to properly separate the (B,C) component
from the radio core, and was therefore discarded. 
The lack of D-array data is very unlikely to result in "filtering out"
any extended CO emission since such emission would have to be extended
over scales $\ga 60\arcsecs$ ($\ga 460$\,kpc).
The fact that we do not find any change in total flux
after including the D-array data verifies this point. 

In total, after calibration overheads, we obtained 6.3\,hr of 
integration time on source for both the
broad and narrow components.  
Calibration and data reduction was done using standard recipes in the NRAO
$\mathcal{AIPS}$ Cookbook.
The amplitude was calibrated with the quasars
3C\,48 and 3C\,286 at the beginning and/or end of each transit. The phase drift
was calibrated using a fast-switching technique in which we observed the nearby
source 04494+63322 every few minutes. The data were obtained in good weather
conditions, and the rms phase fluctuations after calibration was less than
$10\degs$. For the spectral-line mode observations of the narrow component, the
bandpass of the system was calibrated using 0319+415. In all cases the bandpass
calibration averaged over the 7 channels was better than 92\%.
%
%                                                One column figure
%----------------------------------------------------------- S_vib
%______________________________________________________________
Data taken at different times and in different
configurations were combined using DBCON, weighting each dataset with the its
total gridded weight.
%__________________________________________________ One column table
   \begin{table}
      \caption[]{VLA Observations.}
         \label{tbl-1}
     $$
         \begin{array}{p{0.3\linewidth}p{0.3\linewidth}p{0.3\linewidth}}
            \hline
            \noalign{\smallskip}
            Date &  Configuration & $t_{int}$/hrs \\
            \noalign{\smallskip}
            \hline
            \noalign{\smallskip}
            2001 Oct  2  & CD   &  1.4\\
            2001 Oct  9  & CD   &  1.7\\
            2001 Oct 11  & D    &  1.7\\
            2001 Oct 19  & D    &  2.8\\
            2002 Nov 12  & C    &  1.7\\
            2002 Dec 20  & C    &  1.5\\
            \noalign{\smallskip}
            \hline
         \end{array}
     $$
\end{table}
\section{Results \& Analysis}
\begin{figure*}
   \centering
   \includegraphics[width=0.8\hsize,angle=0]{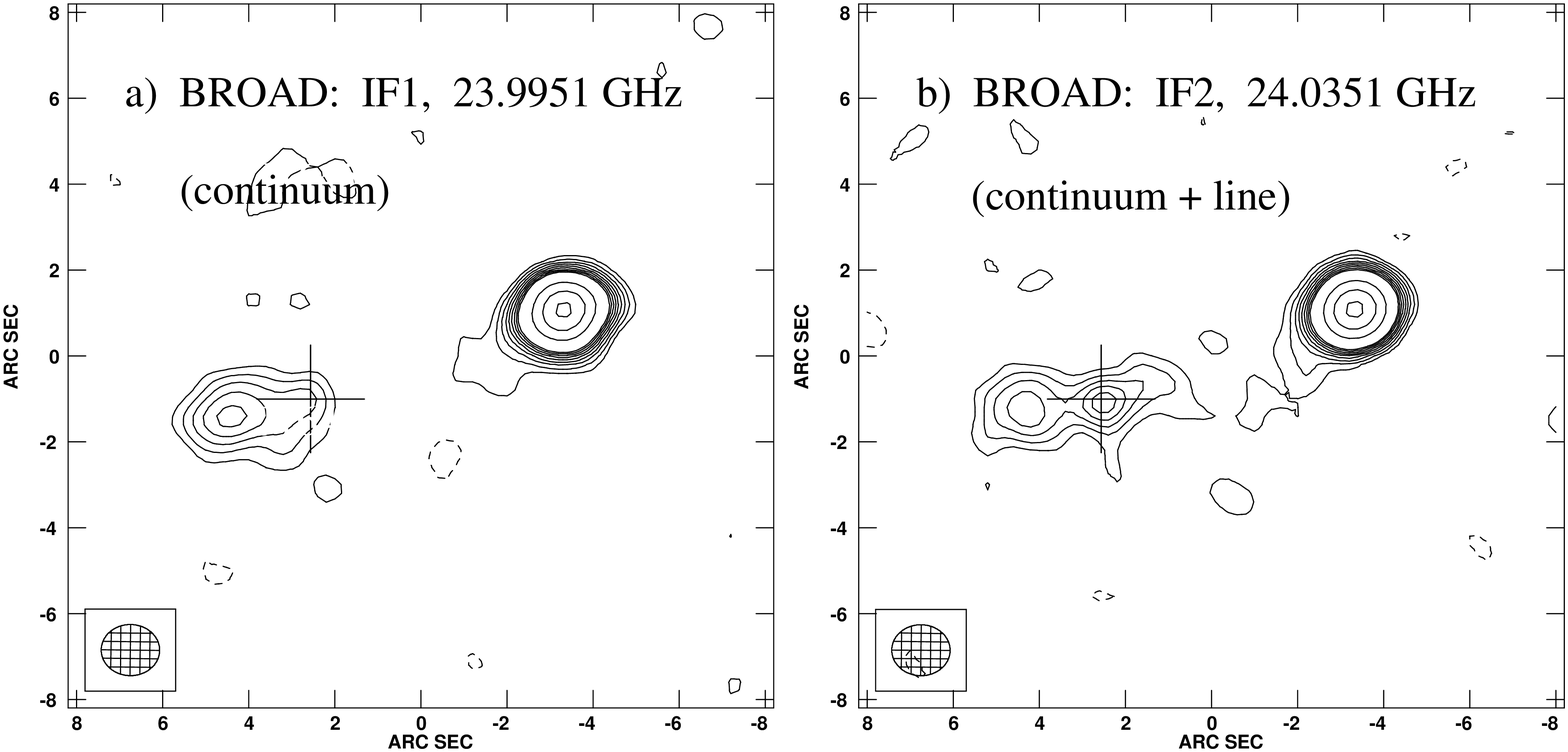}
   \includegraphics[width=0.8\hsize,angle=0]{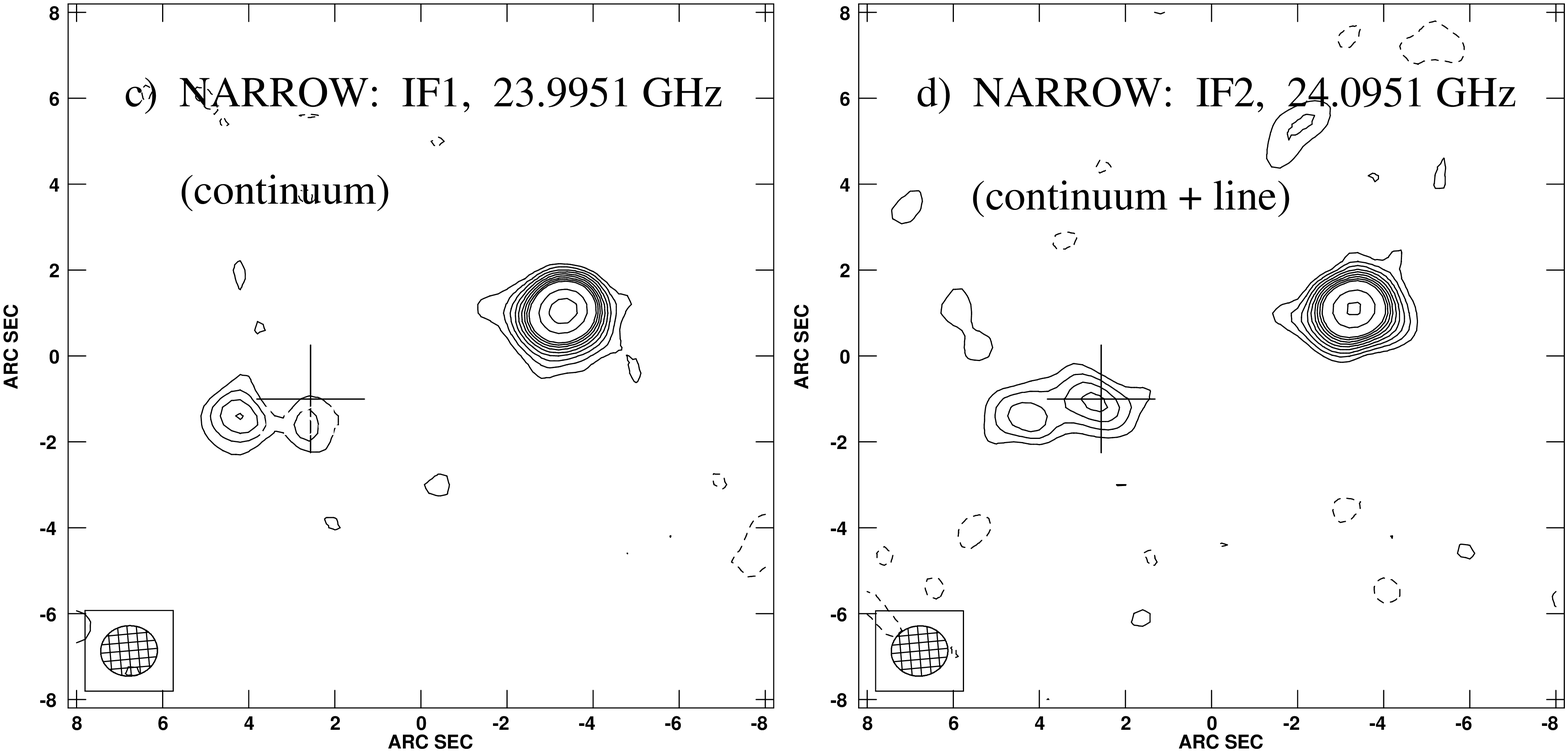}
   \caption{Naturally weighted, untapered maps, showing the morphology of 4C\,60.07 in 
the two IF set-ups. {\bf Top row:} The spatial resolution is $1\parcsec36 \times 1\parcsec19$ with
the major axis P.A.~= $89\degs$. The rms noise is 0.030\,mJy\,beam$^{-1}$. Contours are shown at
-2, 2, 3, 4, 5, 6, 7, 8, 9, 10, 20, 30, and $40 \times 35\,\mu$Jy\,beam$^{-1}$.
{\bf Bottom row:} The spatial resolution (FWHM)
is $1\parcsec32 \times 1\parcsec17$ with the major axis P.A.~= $275\degs$,
and the rms noise level is 0.051\,mJy\,beam$^{-1}$. Contours are shown
at -2, 2, 3, 4, 5, 6, 7, 8, 9, 10, 15, and $20 \times 60\,\mu$Jy\,beam$^{-1}$. 
The cross marks the position of the radio core (R.A.(J2000): 05:12:55.147, DEC.(J2000): +60:30:51.0.}
\label{fig:radio-morphology}
    \end{figure*}

The radio morphology of 4C\,60.7 in the broad and narrow component
IF set-ups is shown in Figure \ref{fig:radio-morphology}.
Figure \ref{fig:radio-morphology}a and b show pure continuum emission and continuum emission plus the
broad CO line emission, respectively; similarly 
for Figure \ref{fig:radio-morphology}c and d but for the narrow component.
%______________________________________________ 
%______________________________________________ 

In Figure \ref{fig:radio-morphology}a, which shows the continuum emission at 23.935\,GHz, we recognise
the two radio lobes seen in Figure \ref{fig:radio-morphology}, although we have failed to separate the
B and C components from the radio core, let alone to 
resolve it into its two sub-components. The emission from the radio core is
clearly stronger and more extended in IF2 which, in addition to the continuum, contains the CO-emission from the
broad component (Figure \ref{fig:radio-morphology}b). 
A similar picture is seen for the narrow component in 
Figure \ref{fig:radio-morphology}c and d. Here, the radio core and B-component are separated in both
IFs.
The (large) negative spectral index ($\alpha \sim -1.7$) of the radio core rules out the possibility
that the excess emission seen in the IF2-maps is due to an increase in continuum emission, 
since the IF2 maps are at a higher frequency than the IF1 maps.
The B and C radio hot spots have even larger negative spectral indices (Carilli et al.~1997)
and can therefore not be the cause of the increased emission either.
This suggests that what we are seeing in the IF2 maps is continuum emission
\emph{plus} \COJ{1}{0} emission from the emission features detected in the \COJ{4}{3} maps.
%______________________________________________ 
   \begin{figure*}
   \centering
   \includegraphics[width=0.4\hsize,angle=0]{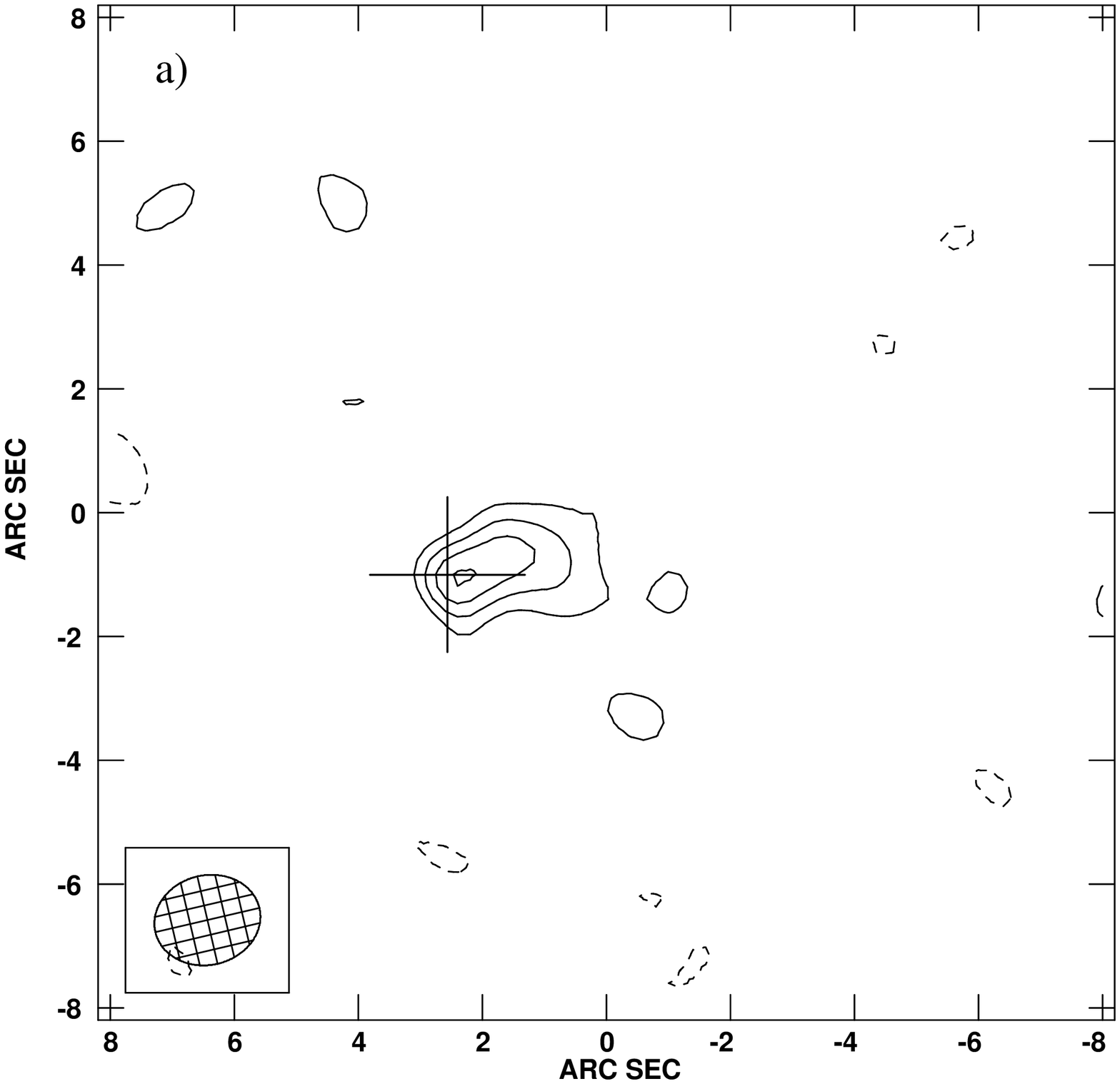}\includegraphics[width=0.4\hsize,angle=0]{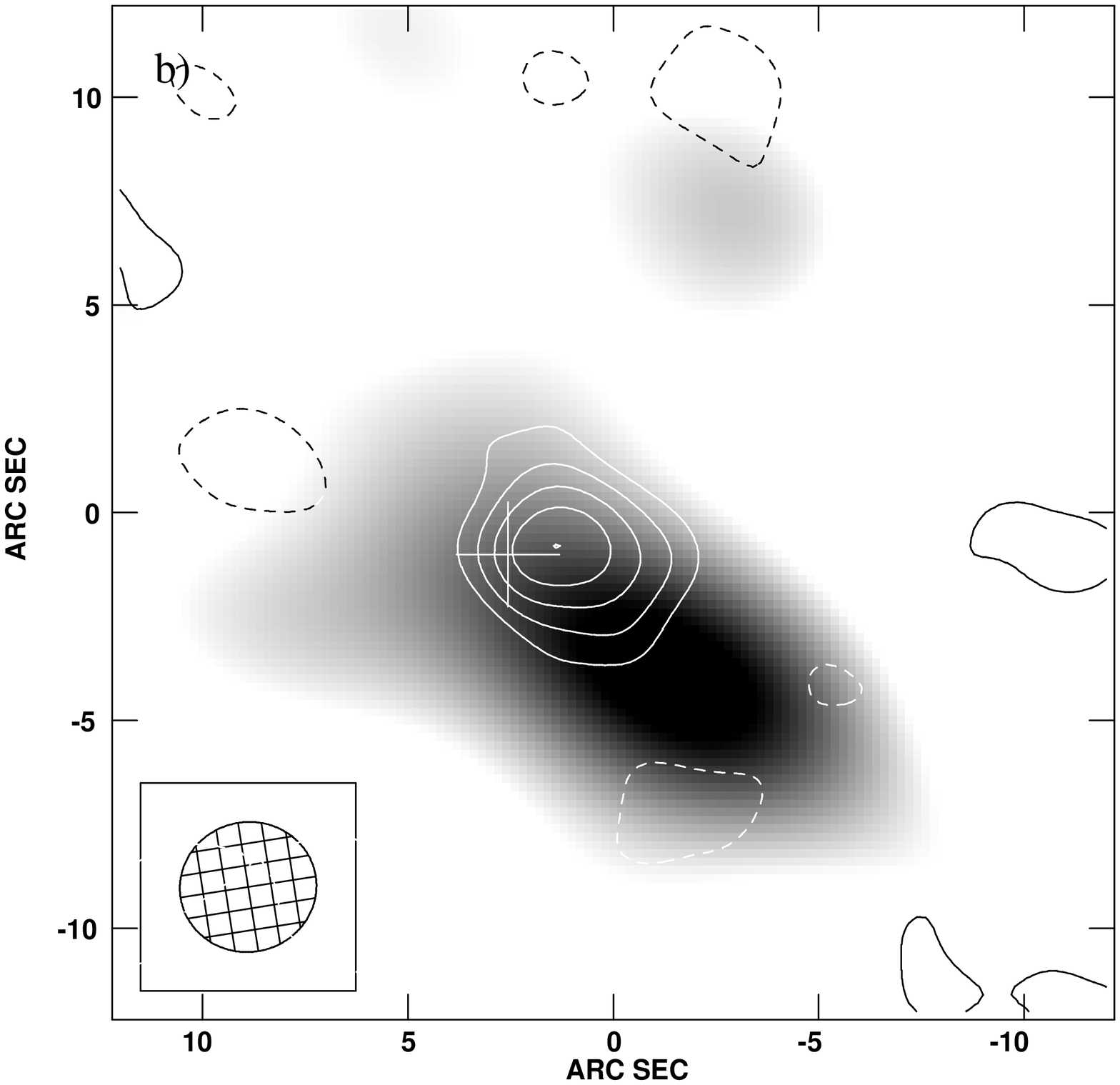}
   \caption{{\bf a)} Naturally weighted \COJ{1}{0} map of the broad component
obtained by combining the entire data-set and tapering
it with a Gaussian with a half-width at 0.30 amplitude of
200k$\lambda$. The spatial resolution is $1\parcsec72 \times
1\parcsec45$ with the major axis P.A.=$283\degs$. The contours are at 
-2, 2, 3, 4, and $5 \times \sigma$, where $\sigma = 35\,\mu$Jy\,beam$^{-1}$. 
{\bf b)} The \COJ{1}{0} emission tapered down to 
60k$\lambda$ ($FWHM = 3\parcsec33 \times 3\parcsec12$) overlaid as contours
on a gray-scale representation of the \J{4}{3} emission by 
Papadopoulos et al.~(2000). The contours are 
-2, 2, 3, 4, 5, and $6\sigma$ with $\sigma = 0.4$\,mJy\,beam$^{-1}$, 
and the gray-scale range is $0.6-2.0$\,mJy\,beam$^{-1}$.}
              \label{fig:CO-broad}
    \end{figure*}
%______________________________________________ 

\subsection{The Broad Line Component}
In order to disentangle the broad CO emission from the continuum, a box was put
around the entire system and a CLEAN-component model of
the continuum emission shown in Figure \ref{fig:radio-morphology}a was constructed. The CLEANing was stopped when
the rms of the residuals reached the noise level in the image. This was then
subtracted in uv-space from the IF2 data using the $\mathcal{AIPS}$ task UVSUB,
resulting in uv-data from which maps free of continuum emission could be
produced.
Figure \ref{fig:CO-broad}a shows the resulting \COJ{1}{0} map of the
broad component. The lack of residual emission at the positions of the
two radio lobes demonstrates the effectiveness of the continuum emission
subtraction.
The \COJ{1}{0} emission is detected at $5\sigma$, peaking at a position which
is coincident with the radio core position but with emission appearing
to extend $2-3\arcsecs$ to the west. 
In comparison, the broad component of the
\COJ{4}{3} emission was found to peak $\sim 4\arcsecs$ south-west of the radio core 
(Papadopoulos et al.~2000).

In Figure \ref{fig:CO-broad}b we have overlaid the \COJ{1}{0} contours on top of a gray-scale image
of the \J{4}{3} emission. The $1-0$ map has been tapered down to 60\,k$\lambda$ in order to 
better match the resolution of the $4-3$ data ($FWHM = 8\parcsec9 \times 5\parcsec5$). 
The offset between the centroids  of the $1-0$ and $4-3$ emission is within 
the positional errors given the large $\J{4}{3}$ synthesized beam, and in general
there is good spatial correspondence between the two.

Assuming that the IF covers the entire line, we can estimate the velocity-integrated
\J{1}{0} flux density using 
\begin{equation}
\int_{\Delta v} S_{\nu_{obs}} dv = \Delta \nu_{IF} \frac{c}{\nu_{obs}} S_{CO},
\end{equation}
where $\Delta \nu_{IF}$ is the width of the IF and $S_{CO}$ is the flux density.
We find $S_{CO}=(0.27 \pm 0.05)$\,mJy which yields a velocity-integrated flux density
of $(0.15\pm 0.03)$\,Jy\,km\,s$^{-1}$, where we have used $\nu_{obs} = 24.035$\,GHz 
and an IF bandwidth of 45\,MHz (560\,km\,s$^{-1}$) instead of 50\,MHz due to bandpass rollover.
The \COJ{1}{0} emission from the broad component
in the uvtaper 200 map appears to be somewhat extended in the north-west direction,
although the low signal-to-noise of the data does not warrant a firm conclusion on this
issue.

\subsection{The Narrow Line Component}
\begin{figure}
   \centering
   \includegraphics[width=1.0\hsize,angle=0]{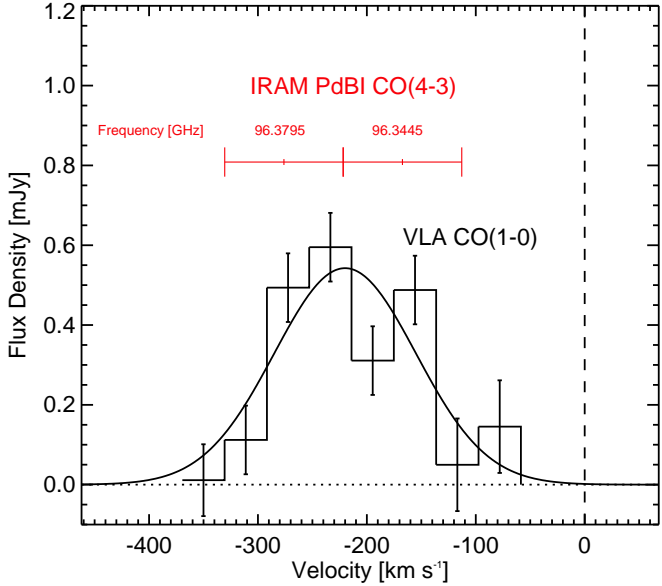}
      \caption{The spectrum of the narrow \COJ{1}{0} component in 4C\,60.07. All channels maps
were tapered with a 200k$\lambda$ Gaussian. The rms noise
in channels 1-2 is 0.12\,mJy\,beam$^{-1}$ at a spatial resolution of $1\parcsec4 \times 1\parcsec3$; 
the noise in channels 3-6 is 0.086\,mJy\,beam$^{-1}$ at a spatial resolution of
$1\parcsec7 \times 1\parcsec4$, and the noise in channels 7-8 is 0.09\,mJy\,beam$^{-1}$
at a spatial resolution of $1\parcsec9 \times 1\parcsec5$. Above the line-profile we
have plotted the velocity coverage of the channels in which 
\J{4}{3} emission was detected (Papadopoulos et al.~2000). The
vertical dashed line corresponds to the systemic velocity ($z=3.788$).}
         \label{fig:CO-narrow}
   \end{figure}

The continuum subtraction in the case of the narrow-line emission
was done in a similar fashion as for the broad component: a
continuum model was constructed by combining all IF1 channels 
and then subtracted from the IF2 channels. Since channels 1 and 2 had
a different uv-coverage than channels 3 to 7 which again had a different
uv-coverage than channels 8 and 9, three different continuum models had to be
constructed, one for each of these three sets of channels. 
In order to increase the signal-to-noise ratio we averaged neighbouring channels
during the imaging of the continuum-subtracted IF2 channels.
We then searched for any residual emission at the expected position
of the narrow component and used the task {\sc imean} to measure
the flux.
The resulting spectrum of the \COJ{1}{0} emission is shown in Figure \ref{fig:CO-narrow}
where we have also plotted the velocity-coverage of the two IRAM PdBI channels in which  
\J{4}{3} emission from the narrow component was detected.
Note that channels 1-2 and 8-9 are somewhat more noisy than the other channels, 
since the corresponding integration time is less. We detect \COJ{1}{0} emission
in channels 3 to 6, which in velocity-space overlaps with the two IRAM PdBI channels
in which the strongest \COJ{4}{3} emission is detected. 
Only very weak \COJ{4}{3} emission was found in the third
IRAM PdBI channel at 96.3095\,GHz which is in agreement with the \J{1}{0} line
profile. From a Gaussian fit to the line profile, we find the \JJ{1}{0} spectrum to
peak at $V \simeq (-220\pm 40)\,$km\,s$^{-1}$ offset from the 
systemic velocity of 4C\,60.07 which 
is consistent with that found for the \JJ{4}{3} line given the large velocity bins. 
The formal linewidth is $\Delta V_{FWHM} \simeq (165 \pm 24)\,$km\,s$^{-1}$, again 
in good agreement with that of the \COJ{4}{3} line profile. 
The fit yields a velocity-integrated flux density of $0.09 \pm 0.01$\,Jy\,km\,s$^{-1}$, see also Table 2.
%----------------------------------------------------------- 
%______________________________________________________________

Combining channels 3 to 6 we obtain the velocity-integrated
\COJ{1}{0} emission map shown in Figure 7a. 
In Figure 7b we have overlaid contours of the \COJ{1}{0} emission 
on a gray-scale image of the \J{4}{3}
emission where the latter has been tapered down to 60\,k$\lambda$. 
The \J{1}{0} emission peaks at $\sim 0\parcsec7$ east of the radio core which is consistent with the
position of the narrow \J{4}{3} component.
Similar to what was found for the \J{4}{3} 
line, in \J{1}{0} the narrow component appears to be more compact and less extended than the broad component.

\section{Discussion}
\subsection{CO Luminosity and Molecular Gas Mass}
The observed \COJ{1}{0} line fluxes for the broad and narrow components in
4C\,60.07 imply intrinsic CO luminosities of $\LOJ{1}{0} = (1.0 \pm 0.2)\times
10^{11}$\,K\,km\,s$^{-1}$\,pc$^2$ and $(6.0 \pm 0.7)\times
10^{10}$\,K\,km\,s$^{-1}$\,pc$^2$, respectively.  For the $4-3$ line
Papadopoulos et al.~(2000) found $\LOJ{4}{3}=(7\pm 1)\times
10^{10}$\,K\,km\,s$^{-1}$\,pc$^2$ and $\LOJ{4}{3}=(3.5\pm 0.8) \times
10^{10}$\,K\,km\,s$^{-1}$\,pc$^2$ for the broad and narrow component,
respectively, where we have computed the luminosities in the cosmology adopted
here. From the \COJ{1}{0} line the molecular gas mass can be found using the
well-known relation $M(\rm{H}_2) = X_{CO} \LOJ{1}{0}$ which relates the \COJ{1}{0}
luminosity with the molecular gas mass (e.g.~Strong et al.~1988). \XCO~is the \COJ{1}{0} 
line luminosity to H$_2$-mass conversion factor which
in the extreme UV-intense environments found in local Ultra Luminous Infra-Red Galaxies,
and presumably also in high redshift galaxies such as 4C\,60.07, 
has a value of about 0.8\,(K\,km\,s$^{-1}$\,pc$^2$)$^{-1} \Msolar$
(Downes \& Solomon~1998).  
In doing so we find molecular gas masses of $M(\rm{H}_2) \sim 8 \times 10^{10}\,\Msolar$ 
and $M(H_2) \sim 5 \times 10^{10}\,\Msolar$ for the
broad and narrow CO emitting components, respectively.
Hence, even for a conservative, non-Galactic value of $X_{\rm{CO}}$
we find that about $\sim 10^{11}\,\Msolar$ of molecular gas is
associated with 4C\,60.07. The estimated gas masses are in very good
agreement with those of Papadopoulos et al.~(2000).

The total gas mass in 4C\,60.07 will of course be larger once the
neutral hydrogen has been accounted for. Assuming a value
of $M(\mbox{HI})/M(\mbox{H}_2) = 2$ which is typically
found in IRAS galaxies (Andreani, Casoli, \& Gerin 1995) we find
a total gas mass of $M_{gas} = M(\mbox{H}_2) + 2M(\mbox{H}_2) = 2.4\times 10^{11}\,\Msolar$
for the broad component and $1.5\times 10^{11}\,\Msolar$ for the narrow component.  
In case that metal-poor gas is also present the gas mass can be even higher.

%This is consistent
%with Papadopoulos et al.~(\cite{Papadopoulos-et-al-2000}) who, using the same conversion
%factor and converted to the same cosmology, estimated the molecular gas masses
%of the broad and narrow to be $\geq 5.5\times 10^{10}\,\Msolar$ and
%$\sim 2.8 \times 10^{10}\,\Msolar$, respectively. These estimates were
%based on an assumed (4-3)/(1-0) ratio of 0.45, and while the two mass estimates actually
%agree reasonably well, the \J{4}{3} estimates yield a factor of two lower masses compared
%to the estimates based directly on the \J{1}{0} line.
%This illustrates the importance of the \JJ{1}{0} line,
%since in the case where only a high-order \COJ{J+1}{J}, $J\geq 2$ line is
%detected one is left with guessing the $r_{J+1,J}$ ratio which may result in
%systematic errors on $M(\rm{H}_2)$ (Papadopoulos et al.~2001; Papadopoulos \& Ivison 2002).
%______________________________________________________________
   \begin{table}
      \caption[]{Observed and Derived Properties for 4C\,60.07.}
     $$
         \begin{array}{p{0.4\linewidth}p{0.25\linewidth}p{0.25\linewidth}}
            \hline
            \noalign{\smallskip}
            Parameter &  4C\,60.07 & 4C\,60.07\\
                      &  (broad)   & (narrow) \\
            \noalign{\smallskip}
            \hline
            \noalign{\smallskip}
            R.A.(J2000)   & 05\hours12\mins54\psec75         &  05\hours12\mins55\psec30\\
            Decl.(J2000)  & +60\degs30\arcmins50\parcsec92   &  +60\degs30\arcmins52\parcsec29\\
            $S_{\rm{CO(1-0)}}$\,[mJy]  &  $(0.27 \pm 0.05)$   & $(0.30 \pm 0.10)$ \\
            $S_{\rm{CO(1-0)}}\Delta V$\,[Jy\,km\,s$^{-1}$]  &  $(0.15 \pm 0.03)$   & $ (0.09 \pm 0.01)$ \\
            \mbox{L$'_{\rm{CO}(1-0)}$}\,[K\,km\,s$^{-1}$\,pc$^2$]  & $(1.0 \pm 0.2) \times 10^{11}$   &  $(6.0 \pm 0.7)\times 10^{10}$\\
            $M(\mbox{H}_2)$ [$\Msolar$]   & $8\times 10^{10}$        &  $5\times 10^{10}$\\
            \noalign{\smallskip}
            \hline
         \end{array}
     $$
\end{table}
%______________________________________________________________
   \begin{figure*}
   \centering
   \includegraphics[width=0.4\hsize,angle=0]{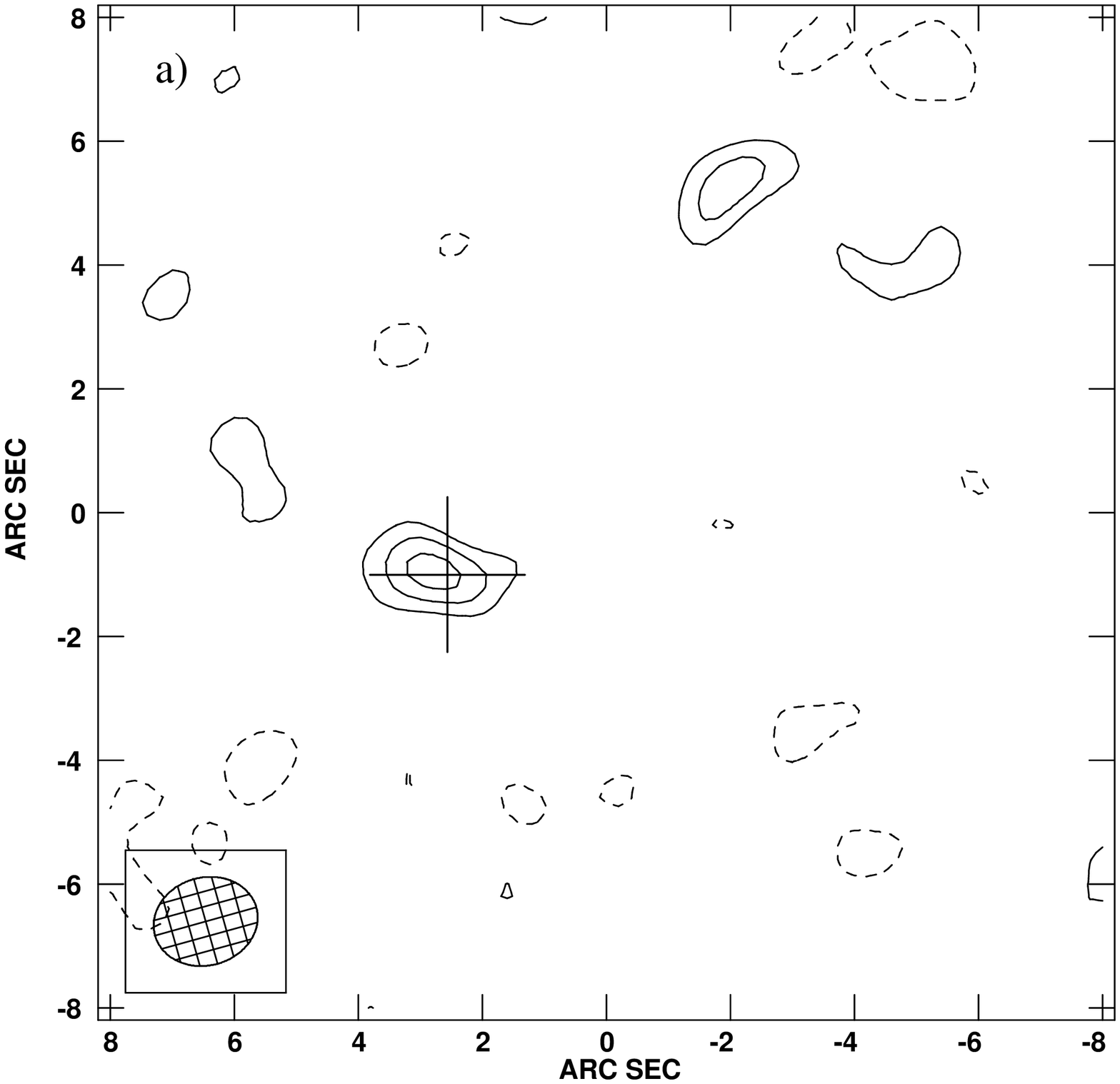}\includegraphics[width=0.4\hsize,angle=0]{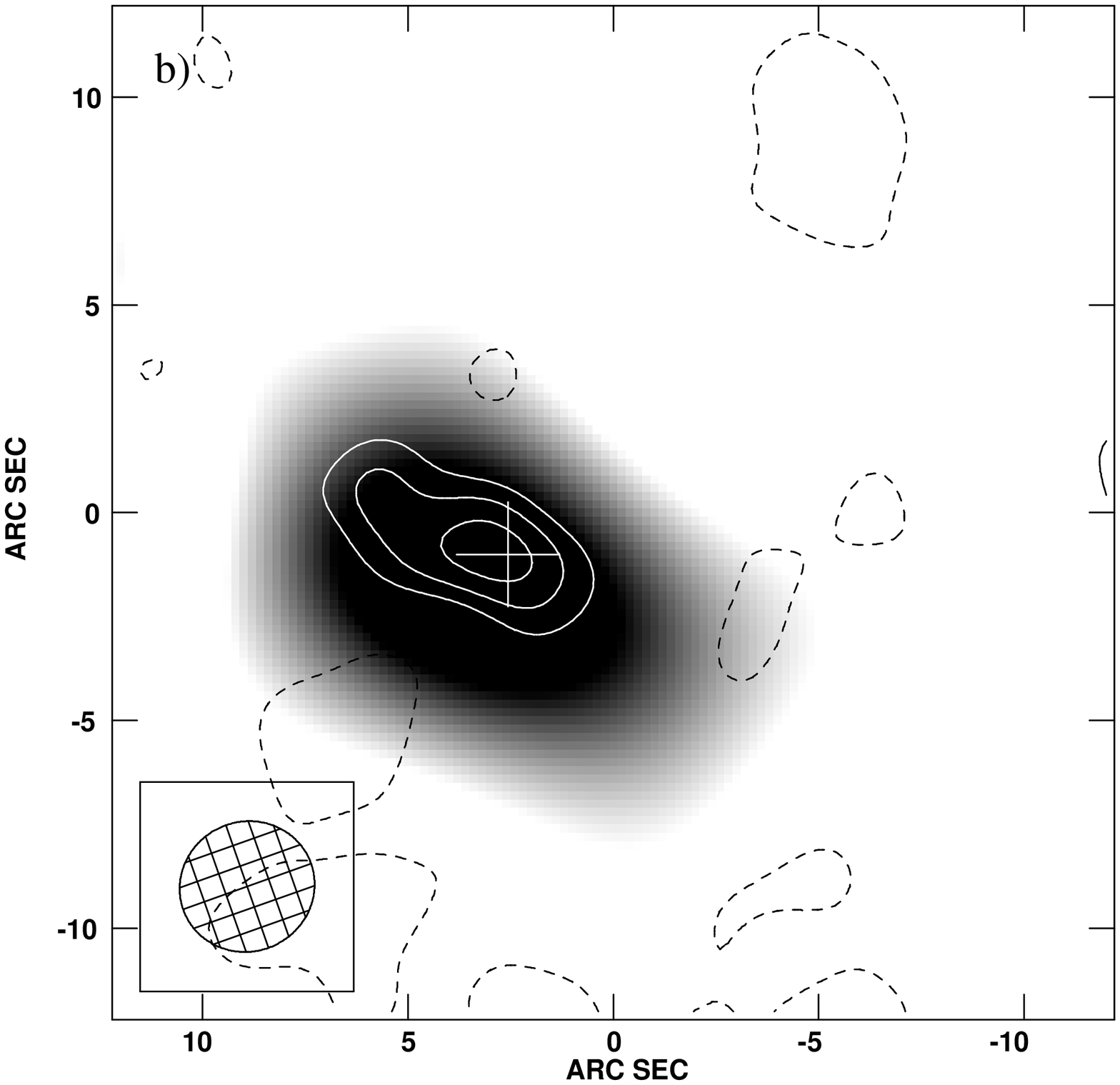}
   \caption{{\bf a)} The narrow component tapered at 200k$\lambda$. The resolution is $1\parcsec72 \times 1\parcsec42$ 
at P.A.~= $285\degs$. Contours are at -2, 2, 3, and $4 \times \sigma$, where $\sigma = 55\,\mu$Jy\,beam$^{-1}$.
{\bf b)} The \COJ{1}{0} emission tapered down to 60k$\lambda$ ($FWHM = 3\parcsec22 \times 3\parcsec07$)
overlaid as contours on a gray-scale representation of the \J{4}{3} detection by 
Papadopoulos et al.~(2000). The contours are -2, 2, 3, and $4\sigma$ with
$\sigma = 0.6$\,mJy\,beam$^{-1}$, and the gray-scale range is $0.6 - 2.0$\,\,mJy\,beam$^{-1}$.}
              \label{Fig2}
    \end{figure*}

\subsection{FIR Luminosity and star-formation efficiency}
In the case where the far-IR luminosity is powered by a starburst and not an AGN,
and all the stellar radiation is absorbed by dust, the
far-IR to CO luminosity ratio, $L_{FIR}/L_{CO}$, provides a rough measure of the 
integrated luminosity of massive stars responsible for heating the dust ($L_{FIR}$)
relative to the amount of fuel available for star formation ($L'_{CO}$). 
We estimate the far-IR luminosity of 4C\,60.07 to be $L_{FIR} \simeq 4\times 10^{13}\,\Lsolar$,
where we have used the 850-$\mu$m flux measurement of Archibald et al.~(2001) and adopted
a dust temperature and spectral index of $T_{d}=50\,$K and $\beta = 1.5$, respectively.
This yields a $L_{FIR}/L_{CO}$ ratio of $\simeq (4\times 10^{13}\Lsolar)/(1.6\times
10^{11}$\,K\,km\,s$^{-1}$\,pc$^2) = 250\,\Lsolar
($\,K\,km\,s$^{-1}$\,pc$^2)^{-1}$, which is similar to values found in ULIRGs
(e.g.~Solomon et al.~1997).
Carilli et al.~(2002a) found continuum-to-line ratios of 350 and 323 in the
QSOs BRI\,1202-0725 ($z=4.70$) and BRI\,1335-0417 ($z=4.41$), respectively.  A somewhat lower value of 200
was found in the $z=4.12$ QSO PSS\,J2322+1944 (Carilli et al.~2002b).  

The current star-formation rate measured in $\Msolar$\,yr$^{-1}$ is given by
\begin{equation}
SFR \simeq (L_{FIR}10^{-10}/\Lsolar) \delta_{MF}\delta_{SB},
\end{equation}
where $\delta_{MF} \sim 0.8-2$ depends on the mass function of the stellar
population, and $\delta_{SB}$ is the fraction of the far-IR luminosity which is
powered by the starburst and not the AGN (Omont et al.~2001). If we 
adopt $\delta_{MF}=0.8$ and conservatively assume that only 50\% of $L_{FIR}$ is powered
by the starburst ($\delta_{SB}=0.5$) we find $SFR \sim 1600\,\Msolar\,\mbox{yr}^{-1}$.
Assuming that the total amount of molecular gas observed towards 4C\,60.07 ends up as fuel for the
starburst, we find that there is enough molecular gas to sustain the inferred star-formation rate 
for $\sim 8 \times 10^7\,\mbox{yr}$.
While this is short compared to the Hubble time it is still sufficient to produce a giant elliptical
with a stellar mass of $M_* \simeq 10^{11}\,\Msolar$.

The efficiency with which stars are being formed, 
i.e.~the rate of star-formation per solar mass of molecular hydrogen, is given 
by $SFE = SFR/M(\mbox{H}_2)$ or equivalently $L_{FIR}/M(\mbox{H}_2)$. 
For 4C\,60.07 we find $SFE = L_{FIR}/M(\mbox{H}_2) \simeq 300\,\Lsolar\,\Msolar^{-1}$, which
is somewhat higher than the $\sim 190\,\Lsolar\,\Msolar^{-1}$ reported by 
Papadopoulos et al.~(2000) who derived their value based on their flux density
measurement of $S_{850\mu m}=11\,$mJy. Here we have used the measurement by Archibald et al.~(2001)
which yields a somewhat higher 850\,$\mu$m flux density of 17\,mJy, although still lower than 
the 22\,mJy reported by Stevens et al.~(2003).

Local ULIRGs exhibit star-formation efficiencies comparable to that of 4C\,60.07,
once the same \XCO-factor has been used (Solomon et al.~1997). 
It is worth stressing that our detection of \COJ{1}{0} enables us to make a
direct comparison of SFEs with that of local ULIRGs since the same gas mass measure
(the \J{1}{0} line) is used for ULIRGs.

The apparently high star-formation efficiencies found for the above systems, could be
severely overestimated if the AGN contributes significantly to the far-IR luminosity.
However, the detection of CO together with the fact that 4C\,60.07 and other HzRGs appear extended on several tens of
kilo-pc scales at submm-wavelengths (Ivison et al.~2000; Stevens et al.~2003) strongly suggests that the far-IR emission
is powered by large-scale starburst and not the AGN. 
Here we must mention that while \COJ{1}{0} may be a good indicator of the total
metal-rich H$_2$ gas resevoir, it may be a poor one regarding the 
dense gas phase ($n \ge 10^5\,\mbox{cm}^{-3}$) that "fuels" star-formation.
The latter could be particular true in the tidally disrupted giant molecular clouds (GMCs) 
expected in ULIRGs
where a diffuse phase may dominate the \COJ{1}{0} emission but has little
to do with star formation. This could be the reason why the $L_{FIR}/L'_{CO}$
ratio is found to be such a strong function of $L_{FIR}$, increasing for the
merger systems associated usually with large far-IR luminosities. Interestingly,
recent work shows that the SFE of dense gas, parametrised by the $L_{FIR}/L_{HCN(1-0)}$
ratio (the HCN \J{1}{0} critical density is $2 \times 10^5\,\mbox{cm}^{-3}$), remains
constant from GMCs all the way to ULIRG system (Gao \& Solomon 2003). 

\subsection{Excitation Conditions}
From the detections of CO in 4C\,60.07 we can infer the CO $(4-3)/(1-0)$ 
velocity/area-averaged brightness temperature 
ratio using $r_{43} = \rm{L}'_{\rm{CO}(4-3)}/\rm{L}'_{\rm{CO}(1-0)}$. The global 
line ratio, i.e.~the line ratio obtained by combining the flux from the two components, is
$r_{43}=0.7^{+0.3}_{-0.2}$. The line ratios for the broad and narrow
components are $0.7^{+0.3}_{-0.2}$ and $0.6^{+0.2}_{-0.2}$, respectively.
Hence, given the large uncertainties we find no significant difference
in the excitation conditions between the two components.
Thus in this case H$_2$ mass estimates solely from \COJ{4}{3} assuming
full thermalisation and optical thickness of the latter (i.e.~$r_{43} \simeq 1$)
would not result in too large errors. However, the sub-thermal excitation of
such high-J CO lines remains a possibility even in starburst environments
(e.g.~Papadopoulos \& Ivison 2002; Greve et al.~2003).

\begin{figure}
   \centering
   \includegraphics[width=1.0\hsize,angle=0]{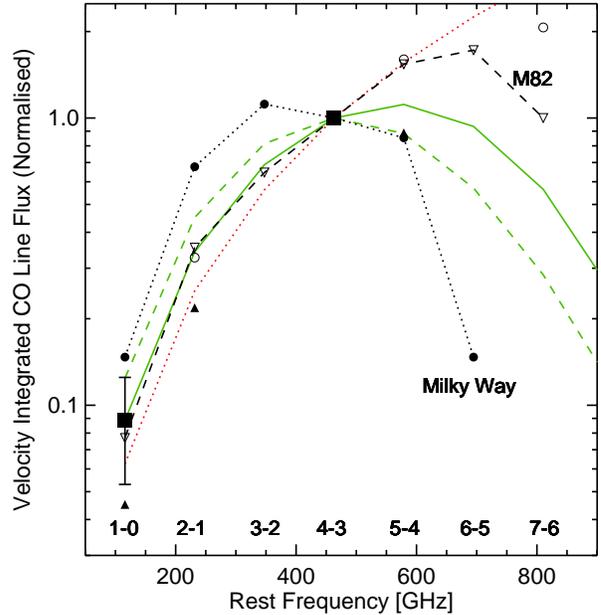} 
      \caption{Velocity-integrated CO line flux densities from
4C\,60.07 are shown as filled squares.
The integrated line fluxes have been normalised to the \J{4}{3} line.
Also shown are the line fluxes
from the $z=4.12$ QSO PSS\,J2322+1944 (filled triangles - Carilli et al.~2002b; Cox et al.~2002),
the $z=4.69$ QSO BRI\,1202-0725 (open circles - Omont et al.~1996; Carilli et al.~2002a),
the local starburst galaxy M\,82 (open triangles - Mao et al.~2000), and
the integrated emission from within the solar radius of the the Milky Way (filled
circles - Fixsen et al.~1999).
The red dotted line shows the line flux increasing as frequency squared which is
expected for optically thick conditions. The green lines show results from
a LVG-model with $T_{kin} = 50\,$K, $X_{\rm{CO}}/(dV/dr) = 
10^{-5}\,$(km\,s$^{-1}$\,pc$^{-1})^{-1}$, and $n(\rm{H}_2)$ equal to $3.0\times 10^3$\,cm$^{-3}$
(solid line) and $2.2\times 10^3$\,cm$^{-3}$ (dashed line).}  
         \label{fig:line_ratio}
   \end{figure}

In Figure \ref{fig:line_ratio} we plotted the velocity-integrated CO line flux
densities (normalised to \J{4}{3}) for the broad and narrow components in 
4C\,60.07 as well as the normalised line fluxes for PSS\,J2322+1944. 
We have used a standard single-component large velocity gradient (LVG)
code to interpret the observed line ratio.
%However, not only is a single line ratio not
%enough to constrain the excitation conditions since severe degeneracy exist between
%parameters such as kinetic temperature and density, but the range of the $r_{43}$ values
%within the errors are so large that they correspond to vastly different conditions.
%Note that since $T_{dust} \simeq T_{kin} = \Delta E_{43}/k$ the $r_{43}$-ratio is mainly
%sensitive to density and optical depth, and 
In fact, given the large range compatible with
our measurements we can only put a lower limit on the density. Indeed, for both
components the upper values for $r_{43}$ are compatible with LTE and optically
thick emission where the ratio is no longer sensitive to the density.

A lower limit on the average gas density can be set by the lowest possible
value of $r_{43}=0.4$ allowed by the observations. Adopting a typical
CO abundance $\Lambda = X_{\rm{CO}}/(dV/dr) =  
10^{-5}\,$(km\,s$^{-1}$\,pc$^{-1})^{-1}$, a 
$T_{CMB} = (1 + z)\times 2.73\,\mbox{K} = 13.07\,$K and a
$T_{kin} = 50\,\mbox{K}$, which is a typical dust temperature in
starburst environments (Colbert et al.~1999; Hughes et al.~1997),
we find $n(\mbox{H}_2) \geq 2 \times 10^{3}\,\mbox{cm}^{-3}$.
Adopting a higher $\Lambda = 10^{-4}$ (unlikely in such kinematically
violent, UV-aggressive environments) lowers the aforementioned limit
to $\sim 600\,\mbox{cm}^{-3}$. A lower assumed temperature of $T_{kin}=30$\,K
does not change these lower limits by much.
For the most likely value of $r_{43}=0.7$ we find 
$n(\rm{H}_2) = 3 \times 10^{3}$\,cm$^{-3}$ ($T_k = 50$\,K, 
$\Lambda  = 10^{-5}\,$(km\,s$^{-1}$\,pc$^{-1})^{-1}$).
%While this model gives the best fit to the 4C\,60.07 data, the lack of 
%high-J transitions precludes us from singling this model out from others.
%This underlines the importance of having many CO lines, spanning as wide as
%range in J-space as possible, when trying to constrain the excitation conditions.

Comparing with the $z=4.12$ QSO PSS\,J2322+1944 which has
a (4-3)/(1-0) line ratio of $1.4\pm0.6$ (Carilli et al.~2002b; Cox et al.~2002)
it appears that the excitation conditions in 4C\,60.07 are less extreme.
Hence, these results may indicate that the molecular gas in 4C\,60.07 is not as dense as
that seen in PSS\,J2322+1944.
%----------------------------------------------------------- 
%______________________________________________________________
Such a comparison, however, is prone to the effects of strong lensing.
While 4C\,60.07 is unlikely to be a strongly lensed system because of its
CO components with widely different line widths, it is not the case for
PSS\,2322+1944 which is known to be lensed (Carilli et al.~2003). 
If the distribution of the \JJ{1}{0} and \J{4}{3} emitting gas
in the source plane differs, differential magnification of the two lines can
result in erroneous intrinsic line ratios.

\subsection{What is the Evolutionary Status of 4C\,60.07?}
A comparison between the molecular gas mass and the dynamical mass
allows for the determination of the evolutionary status of a galaxy, while a comparison
of its dynamical mass with that of present-day spiral or elliptical can point 
toward its possible descendant. Typically, dynamical masses are calculated
assuming the gas is distributed in a disk in Keplerian rotation (e.g.~Carilli et al.~2002a). 
In the case of 4C\,60.07, however, we have sufficient spatial and 
kinematical information to conclude that the two gas components 
do not belong to such a structure. 
The detection of two kinematically distinct gas resevoirs in 4C\,60.07,  
each with a large gas mass, suggest a major merger event.
Hence, a more plausible scenario might be that the two clouds are part of 
spherical system in the process of collapsing. Assuming the system is
virialised, one can apply the virial theorem to derive the following expression for the
dynamical mass 
\begin{equation}
M_{dyn} = 2.33\times 10^9 \left ( \frac{R}{\mbox{kpc}} \right ) 
\left ( \frac{\Delta V_{FWHM}}{100\, \mbox{km\,s}^{-1}} \right ) ^2 \left ( \frac{1}{\alpha (1+q)} \right ),
\label{eq:Mdyn}
\end{equation}
where $q$ is a factor which described the influence of non-gravitational forces
on the virial equilibrium, and $\alpha$ ranges from 0.55 to 2.4 with a typical
value of 1.5 which we shall adopt here (see Bryant \& Scoville (1996) for details). 
A possible caveat to this argument comes from the fact that if the assumption 
of a virialised system is not true, eq.~\ref{eq:Mdyn} will tend to overestimate
the true dynamical mass.
Adopting a radius of the sphere corresponding to half the
separation between the two components, i.e. $R\sim 15$\,kpc (corresponding
to $2\arcsecs$), and a FWHM width equal
to the velocity offset between the line centers of the two gas resevoirs, 
i.e.~$\Delta V_{FWHM}\sim 700$\,km\,s$^{-1}$, leads to an enclosed
dynamical mass of $M_{dyn}\la 1.1\times 10^{12}\,\Msolar$,
where we have assumed that non-gravitational forces are negligible, i.e.~$q << 1$.
Note that since Bryant \& Scoville (1996) derived eq.~\ref{eq:Mdyn} by examining
a large number of clouds filling up a spherical region, we have in the
above assumed that the velocity and spatial separation between the
two CO components is a fair approximation for $\Delta V_{FWHM}$ and $R$ in
eq.~\ref{eq:Mdyn}. Alternatively, we can do the more simple calculation of
finding the virial mass required to keep the two clouds bound by a spherical
potential. In this case the velocity dispersion becomes $\Delta V = 
(1/2) 700\,\mbox{km\,s}^{-1} = 350\,\mbox{km\,s}^{-1}$, and we find 
$M_{virial} = G^{-1} R \Delta V^2 \simeq
4.3 \times 10^{11}\,\Msolar$, where we have assumed $R=15$\,kpc.
Projection effects will only tend to make $R$ and $\Delta V_{FWHM}$, and 
therefore also the above mass estimates, larger. 
Thus, 4C\,60.07 is a massive system with a dynamical
mass comparable to that of giant elliptical galaxies
seen in the present day Universe. 
The inferred ratio of the molecular-to-dynamic mass for
the system as a whole is $M(\mbox{H}_2)/M_{dyn} \sim 0.12$, 
and the total gas fraction is $M_{gas}/M_{dyn} \sim 0.35$.
The high gas fraction, along with the large velocity dispersion of the system
suggest that 4C\,60.07 is a very young system in its early stages of formation.

The fact that such a massive ($M_{dyn} \sim 10^{12}\,\Msolar$) system
has assembled at $z=3.8$, which corresponds to a time when the Universe
was only $\sim 10$\% of its present age, seems to favour the monolithic collapse
model (e.g.~Eggen, Lynden-Bell \& Sandage 1962; Tinsley \& Gunn 1976) 
of massive ellipticals over the hierarchical formation scenario 
(e.g.~Baron \& White 1987; Baugh et al.~1996; Kauffmann \& Charlot 1998).
Further in support of the monolithic collapse scenario is the large star-formation rate 
($SFR\sim 1600\,\Msolar$yr$^{-1}$) we infer
from the far-IR luminosity. If such a large star-formation rate can
be sustained it is capable of producing a giant elliptical in a time scale
comparable to the dynamical time. The dynamical time-scale for 
a system with a mass $M$ within a radius $R$  is
\begin{equation}
t_{dyn} = 7.42 \times 10^{11} \left ( \frac{R}{\mbox{kpc}} \right ) ^{3/2} \left ( \frac{M}{\Msolar} \right )^{-1/2},
\end{equation}
where $t_{dyn}$ is measured in years (see p.~37 of Binney \& Tremaine 1987). For a giant
elliptical with a mass $M_{ell} = 10^{12}\,\Msolar$ and a radius $R=30$\,kpc, we find
$t_{dyn} \simeq 1.2 \times 10^8$\,yr. Hence, in order to form a giant elliptical
within a dynamical time scale, star-formation rates of the order $M_{ell}/t_{dyn} 
\simeq 8000\,\Msolar\,$yr$^{-1}$ are required,
which is comparable to what we find in 4C\,60.07. 
It is important to note that even if the true dynamical mass or the
star-formation rate of this system are less than the above estimates, the fact that
at $z=3.8$ about $1.3 \times 10^{11}\,\Msolar$ in H$_2$ gas mass alone has assembled
within 15\,kpc points towards a major galaxy forming merger rather than a gradual 
accumulation of mass over time and redshift. This mass can only be revised 
upwards and typically reach $\sim 4 \times 10^{11}\,\Msolar$ if HI is also 
accounted for, thereby making the case for a monolithic collapse scenario even stronger. 

4C\,60.07 is not the only system with extended CO emission.
Molecular gas distributed on tens of kpc scales, sometimes 
in separate gas components, have been observed in high
redshifts quasars (e.g.~Carilli et al.~2002a; Papadopoulos et al.~2001)
and radio galaxies (De Breuck et al.~2003; De Breuck et al.~in prep.).
%.  In the quasar
%BRI\,1202-0725 ($z=4.695$) for example, Carilli et al.~(2002a) found the molecular
%gas to consists of a narrow and a broad-line component.
%In addition Carilli et al.~(2002a) 
%found that the southern component is made up of two
%roughly equal sources, separated by $0\parcsec3$, and thereby demonstrated the
%importance of high-resolution observations if one wants to map the true
%distribution of the molecular gas. Similar examples are BRI\,1335-0417 which
%also seems to be comprised of at least two components $1\parcsec3$ apart
%(Carilli et al.~2002a), and the $z=3.087$ radio galaxy B3\,J2330+3927
%which is resolved in CO (De Breuck et al.~2003a).
Thus, there is evidence to show that at least some
luminous active galaxies in the early Universe,
in addition to harbouring supermassive black holes in their centres, are
associated with massive reservoirs of molecular gas distributed
on tens of kpc scales which are in the process of merging. 
This observed coevality between large scale mergers
of gaseous subsystems and the epoch of
AGN-activity might provide clues to the origin of the tight relationship observed locally
between the velocity dispersion of spheroids and the mass of their central
black holes (Ferrarese \& Merritt 2000).

\begin{acknowledgements}
TRG acknowledges support from the Danish Research Council and from the EU RTN
Network POE. PPP acknowledges a Marie Curie Individual Fellowship 
HPMT-CT-2000-00875. We are grateful to Ignas Snellen and Philip Best for useful
advice on $\mathcal{AIPS}$. We are also greatly indebted to Carlos De Breuck and Michiel Reuland
for providing us with the IRAM PdBI \COJ{4}{3} data. 
\end{acknowledgements}

\end{document}